\documentstyle[aps,preprint,epsf]{revtex}
\firstfigfalse
\newcommand{\bea}{\begin{eqnarray}}
\newcommand{\beal}[1]{\begin{eqnarray}\label{#1}}
\newcommand{\eea}{\end{eqnarray}}
\newcommand{\ben}{\begin{eqnarray*}}
\newcommand{\een}{\end{eqnarray*}}
\newcommand{\beq}{\begin{equation}}
\newcommand{\eeq}{\end{equation}}
\newcommand{\be}{\begin{equation}}
\newcommand{\bel}[1]{\begin{equation}\label{#1}}
\newcommand{\ee}{\end{equation}}

\newcommand{\gl}[1]{(\ref{#1})}

\renewcommand{\a}{\alpha}
\renewcommand{\b}{\beta}
\renewcommand{\d}{\delta}

\newcommand{\s}{\sigma}
\renewcommand{\l}{\lambda}

\newcommand{\G}{\Gamma}

\newcommand{\dvc}[1]{\stackrel
                  {\lower1mm\hbox{$\scriptscriptstyle\!\leftharpoonup
                   \!\!\!\!\!\rightharpoonup$}}{#1}}

\newcommand{\0}{\over }

\newcommand{\2}{\frac{1}{2}}

\def\({\left(}   \def\){\right)}

\def\gtrsim{\mbox{\,\raisebox{.3ex}{
           $>$}$\!\!\!\!\!$\raisebox{-.9ex}{$\sim$}\,\,}}

\preprint{{\tt gr-qc/9510038}  /  DESY 95-190 }
\begin{document}
\tighten
\title{Quantum Field Theory at Finite Temperature\\ and
Cosmological Perturbations\thanks{Talk presented at the Third
   Workshop on Quantum Field Theory under the Influence of External Conditions,
   Leipzig, Germany, 18-22 September, 1995}
}
\author{Anton K. Rebhan\thanks{On leave of absence from
        Institut f\"ur Theoretische Physik der
         Technischen Universit\"at Wien,
Wiedner Hauptstr.~8--10, A-1040 Wien, Austria;
email: rebhana@email.tuwien.ac.at}
         }
\address{DESY, Gruppe Theorie,\\
        Notkestra\ss e 85, D-22603 Hamburg, Germany}

\maketitle

\begin{abstract}
It is shown how quantum field theory at finite temperature
can be used to set up self-consistent and gauge invariant
equations for cosmological perturbations sustained by an
ultrarelativistic plasma. While in the collisionless case,
the results are equivalent to those obtained from the
Einstein-Vlasov equations, weak self-interactions in the
plasma turn out to require the full machinery of perturbative
thermal field theories such as resummation of hard thermal loops.
Nevertheless it is still possible to use the same methods that yielded
exact solutions in the collisionless case.
\end{abstract}


\narrowtext
\bigskip\bigskip

In order to account for the present large-scale structure of the
universe such as galaxies, clusters, superclusters, voids, etc., \cite{Boerner}
a cosmological model built on homogeneous and isotropic geometries
requires certain imperfections in its symmetries. Through the
universally attractive nature of gravitations, initially small
perturbations can grow, in particular
once that the universe becomes matter dominated and the
(maximal) pressure provided by radiation has become inoperative.
This picture, which is based on a big-bang scenario, has found
dramatic support by the search for and discovery of tiny anisotropies
in the cosmic micro-wave background, which in a Friedmann-Robertson-Walker
(FRW)
universe have wavelengths far exceeding the size of the Hubble horizon
at the time when this radiation decoupled from the primordial matter
\cite{Smoot}.

Whatever the origin of these small deviations from homogeneity and isotropy,
there is a rather long epoch of radiation domination which is thought
to be well described by a nearly perfect FRW model with metric
perturbations evolving in a linear regime. The basic equations for
these ``cosmological perturbations'' are nothing else than
the perturbed Einstein equations,
\bel{dGmn}
\d G^{\mu\nu}\equiv{\d(R^{\mu\nu}-\2 g^{\mu\nu}R)\0\d g^{\a\b}}
\d g^{\a\b}=-8\pi G\,\d T^{\mu\nu}.
\ee

In order to have a close set of equations, these have to be
supplied with information on the response $\d T^{\mu\nu}$ of
the energy-momentum tensor to metric perturbations $\d g^{\a\b}$.
In a hydrodynamic approach, this is done by sufficiently restricting
the form of $\d T^{\mu\nu}$, specifying the equations of state,
and imposing covariant conversation of the full energy-momentum
tensor in the perturbed geometry. The simplest case is the one
of a perfect (radiation) fluid, which has been studied in the
pioneering work of Lifshitz \cite{Lifshitz}. Many generalizations
have since been worked out, and have been cast into a gauge invariant
form by Bardeen \cite{Bardeen}. A modern geometrical justification and
generalization has been given recently by Ellis and co-authors \cite{EB}.

A more fundamental description of the behaviour of the primordial
matter, which in the early universe is mostly a hot plasma of
elementary particles, is
usually implemented through kinetic theory \cite{Kin}. However, a
truly fundamental description eventually has to take into account
quantum field theory. In the following I shall show that
an interesting part of the theory of cosmological perturbations can
be investigated through the techniques developed for quantum field
theory at finite temperature \cite{Kapusta},
namely the case of a weakly interacting
ultrarelativistic particle plasma.  In the limiting case of a
collisionless ultrarelativistic plasma it turns out to be even possible
to obtain exact analytic results \cite{KR}
where only numerical ones where
known before; in the case of weak self-interactions one can still
find analytic results \cite{Rapid} which involve such issues as resummation
of hard thermal loops that would be very difficult to include in
a (quantum) kinetic approach.

For temperatures $T\ll m_{\mathrm Planck}$ it is sufficient to
treat the gravitational field as a classical background field.
The energy-momentum tensor can then be defined by the one-point
function
\bel{Tmn}
T_{\mu\nu}(x)={2\0\sqrt{-g}}{\d\G[g]\0\d g^{\mu\nu}},
\ee
where $\G[g]$ is the effective action functional that contains all
the contributions besides the classical Einstein-Hilbert action.
When derived from this effective action, covariant conservation of
the energy-momentum tensor is automatic and need not be imposed
as a constraint.

The response under perturbations in the metric field is given by
\begin{equation}
\delta T_{\mu\nu}(x)=\int d^4y
{\delta T_{\mu\nu}(x)\over \delta g^{\alpha\beta}(y)}
\delta g^{\alpha\beta}(y).
\end{equation}
Hence, $\delta T_{\mu\nu}$ is determined by the gravitational
polarization tensor (or ``thermal graviton self-energy'')
\begin{equation}
\Pi_{\mu\nu\alpha\beta}(x,y)\equiv
{\delta^2\Gamma\over\delta g^{\mu\nu}(x)
\delta g^{\alpha\beta}(y)}=\frac12{\delta(\sqrt{-g}T_{\mu\nu}(x))\over
\delta g^{\alpha\beta}(y)}.
\label{pi}
\end{equation}

In particle physics terminology, \gl{Tmn} and \gl{pi} are the sets of
one-particle irreducible diagrams with one and two external graviton line(s)
in the background field $g_{\mu\nu}$ given by the cosmological
model on which one wants to study the dynamics of cosmological
perturbations. The concept of thermal equilibrium makes rigorous
sense in conformally trivial situations where $g_{\mu\nu}(x)=
\sigma(x)\eta_{\mu\nu}$. This is indeed the case
with almost all of the cosmological models of interest.
If one knows how the effective action
transforms under conformal rescalings of the metric, then the
entire problem of determining the
highly nonlocal function \gl{pi} (and thus the response of
the plasma) can be reduced to the evaluation of Feynman diagrams
in flat space, where momentum-space techniques can be used.
In flat space, temperature can be introduced through periodicity
in imaginary time, and retarded Green functions in real time
are obtained by analytic continuation.

In the high-temperature limit,
where all the momenta and masses of the internal
particles are assumed to be
much smaller than temperature, the effective action in fact turns
out to be invariant under conformal rescalings, so \gl{pi} on a
curved space with vanishing conformal Weyl tensor
can be reconstructed by the simple transformation
\begin{eqnarray}
\label{ftpi}
\left. \Pi_{\mu\nu\rho\sigma}(x,y)
\right|_{g_{\mu\nu}=\sigma\eta_{\mu\nu}} &&  \nonumber \\
= \sigma(x)
\int  {d^4 k \over (2\pi)^4}  & &
e^{i k (x - y)} \left.
\tilde{\Pi}_{\mu\nu\rho\sigma}(k)\right|_{\eta} \sigma(y) \ .
\end{eqnarray}

The Planck mass, which we assumed to be much larger than temperature,
does not explicitly appear in $\tilde\Pi$ since we are treating
the metric field as classical and no higher loop diagrams
with graviton self-interactions are involved. So we only need to assume
that the (zero-temperature)
masses of the thermal matter are small compared to
temperature (i.e., the plasma is ultrarelativistic), and that
the relevant momentum scales are likewise so. Fortunately, this
is just the case of interest with cosmological perturbations.
If the latter have typical wavelengths of the order of the Hubble
horizon, then $k/T\sim \sqrt{GT^2} \propto T/m_{\mathrm Planck}\ll1$.

One-loop diagrams correspond to collisionless thermal matter which
has only gravitational interactions. The leading temperature
contributions to \gl{pi} have been first calculated in \cite{Rebhan91}
(see also Ref.~\cite{ABFT})
and turn out to have a universal structure, where only the
overall factor varies among the various forms of thermal matter
according to their energy density. It is highly nonlocal and
comes with a complicated tensor structure, since with
$\eta^{\mu\nu}$, $u_\mu=\delta^0_\mu$, and $K^\mu=(K^0,{\bf k})$,
one can build 14 tensors to form a basis
for $\tilde\Pi^{\mu\nu\alpha\beta}(K) = \rho \sum_{i=1}^{14} c_i(K)
T_i^{\mu\nu\alpha\beta}(K)$, see Table \ref{table1}.

However, $\tilde\Pi$ satisfies the Ward identities corresponding to
diffeomorphism invariance and conformal invariance, and this
reduces the number of independent structure functions to 3.
They can be chosen as
\begin{equation}
A(K)\equiv\tilde\Pi_{0000}(K)/\rho,\quad
B(K)\equiv\tilde\Pi_{0\mu}{}^\mu{}_0(K)/\rho,\quad
C(K)\equiv\tilde\Pi_{\mu\nu}{}^{\mu\nu}(K)/\rho
\end{equation}
($\rho = T_{00}$), and
the $c_{1\ldots14}$ are determined by the linear combinations given
in Table \ref{table2}.

The universal result for ultrarelativistic collisionless thermal matter
then reads
\begin{equation}\label{ABC1}
A^{(1)}(K)=\omega\,{\rm artanh}{1\over \omega}-\frac54,\qquad
B^{(1)}=-1,\qquad C^{(1)}=0,
\end{equation}
with $\omega\equiv K_0/k$.

Cosmological perturbations can be classified according to their
transformation behaviour under spatial coordinate transformations
\cite{Bardeen}
as scalar, vector, or tensor, which corresponds to compressional,
rotational, or radiative perturbations in the plasma. The above
3 independent components of $\tilde\Pi$ determine, in certain
combinations, the connection between the respective perturbations
in the energy-momentum tensor and in the metric field.

In the radiation-dominated epoch the standard choice is that of
a spatially flat Einstein-de Sitter model with
line element
\bel{ds2}
ds^2=\sigma(\tau)(d\tau^2-d{\bf x}^2),\qquad
\sigma(\tau)={8\pi G\rho_0\03}\tau^2,
\ee
($\rho_0$ is the energy density when $\sigma=1$),
and, given \gl{ds2}, it is
moreover natural to decompose all perturbations in plane waves,
since in linear perturbation theory the different modes evolve
independently. The problem is thus reduced to a one-dimensional
one, and it is convenient to introduce a dimensionless time variable
\bel{xdef}
x\equiv{k\tau}={R_H\0\lambda/(2\pi)},
\ee
which measures the (growing) size of the Hubble horizon over
the wavelength of a given mode (which is constant in comoving coordinates).

Of all the numerous components of \gl{dGmn}, only a few are
independent by virtue of general covariance and turn out to
involve only those gauge-invariant combinations of the components of
the metric perturbations $\d g_{\mu\nu}$ that have been
studied by Bardeen \cite{Bardeen}. For instance, the scalar part of
metric perturbations can be parametrized in terms of four
scalar functions
\be
\d g_{\mu\nu}^{(S)}=\s(\tau)
\pmatrix{C & D_{,i} \cr D_{,j} & A\d_{ij}+B_{,ij}\cr}
\ee
of which always two can be gauged away. Instead of fixing a gauge,
we can also use the gauge-invariant combinations
\bea
\Phi&=& A+{\dot\s\0\s}(D-\2\dot B) \\
\Pi &=& \2(\ddot B+{\dot\s\0\s}\dot B+C-A)-\dot D-{\dot\s\0\s}D,
\eea
where a dot denotes differentiation with respect to the conformal time
variable $\tau$.

Each spatial Fourier mode with wave vector $\bf k$ is
related to perturbations in the energy density and anisotropic pressure
according to
\be
\delta={1\03}x^2 \Phi,\qquad \pi_{anis.}={1\03}x^2 \Pi.
\ee
Here energy density perturbations $\d$ are defined with respect to
space-like hypersurfaces representing everywhere the local rest frame
of the full energy-momentum tensor, whereas $\pi_{anis.}$ is an
unambiguous quantity, since there is no anisotropic pressure in
the background.

Correspondingly, when specifying to scalar perturbations, there are
just two independent equations contained in \gl{dGmn}. Because
of conformal invariance, the trace of \gl{dGmn} is particularly
simple and yields a finite-order differential equation in $x$,
\be
\Phi''+{4\0x}\Phi'+{1\03}\Phi={2\03}\Pi-{2\0x}\Pi'
\ee
(a prime denotes differentiation with respect to the dimensionless
time variable $x$).
The other components, however, involve the nonlocalities of
the gravitational polarization tensor. These lead to an
integro-differential equation, which upon imposing retarded boundary
conditions reads \cite{KR}
\bel{scp}
(x^2-3)\Phi+3x\Phi'=6\Pi-12\int_{x_0}^x dx'\,
j_0(x-x')(\Phi'(x')+\Pi'(x'))+\varphi(x-x_0)
\ee
where $j_0(x)=\sin(x)/x$ arises as Fourier transform of
$A(\omega)$ in \gl{ABC1}.
$\varphi(x-x_0)$ encodes the initial conditions,
the simplest choice of which corresponds to $\varphi(x-x_0)\propto j_0(x-x_0)$.

Similar integro-differential equations have been obtained
from coupled Einstein-Vlasov equations in particular gauges, and
the above one can be shown to arise from a gauge-invariant
reformulation of classical kinetic theory \cite{AKRDS}.
Usually, these equations were studied numerically, but in fact
they can be solved analytically \cite{KR}. If initial conditions
are formulated for $x_0\to0$, a power series ansatz for $\Phi$ and $\Pi$
leads to recursion relations that can be solved and lead to an
alternating series
that converges faster than trigonometric functions.

This also holds true for the vector and tensor perturbations and
when the more realistic case of a two-component system of perfect
radiation fluid and ultrarelativistic plasma is considered \cite{Rebhan92}.

\begin{figure}
\centerline{ \epsfxsize=5in
\epsfbox{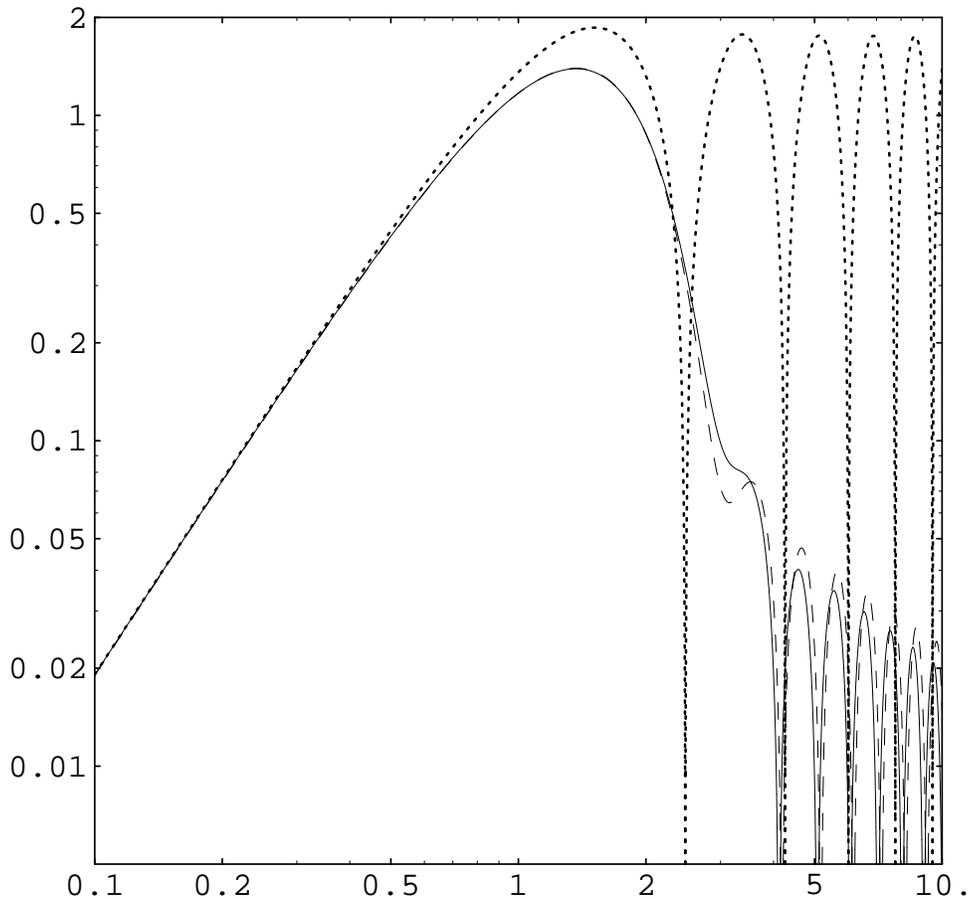} }
\caption{The energy-density contrast (arbitrary
normalization) as a function of $x/\pi$ for a collisionless
ultrarelativistic plasma (full line), a scalar plasma with
quartic self-interactions $\lambda\phi^4$ and $\lambda=1$ (dashed
line), and a perfect radiation fluid (dotted line). }
\label{plot}\end{figure}

In Fig.~\ref{plot}, the solution for the energy-density contrast
is given in a doubly-logarithmic plot (full line) and compared with
the perfect-fluid case (dotted line). In the latter, one has growth
of the energy-density contrast as long as the wavelength of the
perturbation exceeds the size of the Hubble radius ($x\ll1$).
After the Hubble horizon has grown such as to encompass about
one half wavelength ($x=\pi$), further growth of the perturbation
is stopped by the strong radiation pressure, turning it into
an (undamped) acoustic wave propagating with the speed of sound in radiation,
$v=1/\sqrt3$. The collisionless case is similar as concerns
the superhorizon-sized perturbations, but after horizon crossing,
there is strong damping $\sim1/x$, and the phase velocity is about 1.
This indeed reproduces the findings of the numerical studies
of Ref.~\cite{Bond}. They can be understood as follows: a energy-density
perturbation consisting of collisionless particles propagates with
the speed of their constituents, which in the ultrarelativistic case
is the speed of light, and there is collisionless damping
in the form of directional dispersion.

While with purely collisionless ultrarelativistic matter, all results
are equivalent \cite{AKRDS}
to solving the classical Einstein-Vlasov equations,
a quantum-field-theoretical treatment comes into its own when
self-interactions within the thermal matter are taken into account.
In a kinetic treatment one could add in a collision term to the
coupled Einstein-Boltzmann equations, but eventually one would
have to abandon the classical concept of a distribution function
for the thermal matter. A virtue of the above thermal-field-theoretical
approach is that everything is formulated in purely geometrical terms,
without explicit recourse to perturbations in the (gauge variant)
distribution function.

\begin{figure}
\centerline{ \epsfxsize=3.5in
\epsfbox[82.9728 357.2 385.027 488.732]{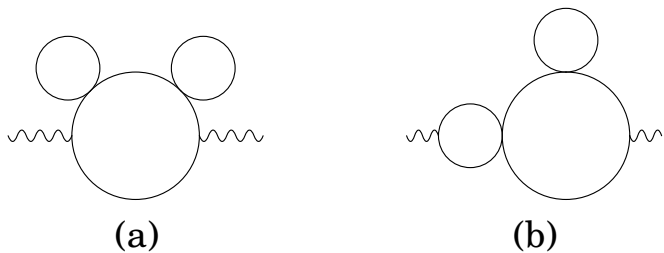} }
\caption{Two examples of infrared divergent graphs beyond two-loop order.}
\label{f3}\end{figure}

In Ref.~\cite{Nrs2}, the gravitational polarization tensor has been
calculated in a $\lambda\phi^4$ theory through order $\lambda^{3/2}$.
The next-to-leading order contributions to $\Pi_{\mu\nu\a\b}$
at order $\lambda^1$ are contained
in the high-temperature limit of two-loop diagrams and their evaluation is
straightforward. However, starting at three-loop order, there are
infrared divergences which signal a breakdown of the convential
perturbative series. This is caused by the generation of a thermal
mass $\propto \sqrt{\lambda}T$ for
the hot scalars. If this is not resummed into a
correspondingly massive scalar propagator, repeated insertions
of scalar self-energy diagrams in a scalar line produces arbitrarily
high powers of massless scalar propagators all with the same momentum,
and thus increasingly singular infrared behaviour (Fig.~\ref{f3}a).

However, it is not sufficient to resum this thermal mass for the
hot scalars. After all, this would break conformal invariance.
Indeed, there are also vertex subdiagrams $\propto\l T^2$
that have a similar
effect as a self-energy insertion, see. Fig.~\ref{f3}b.
As in the hard-thermal-loop
resummation program developed for high-temperature
quantum chromodynamics \cite{BP}, one has to resum also
nonlocal vertex contributions.
Doing so, the result turns out to satisfy both the diffeomorphism
and conformal Ward identities.

In the low-momentum limit that is of interest in our application to
cosmological perturbations, the function $A$ in \gl{ABC1} that
governs the evolution of scalar perturbations reads through
order $\lambda^{3/2}$
\begin{eqnarray}\label{Ar}
A &=& \omega\,{\rm artanh}{1\over \omega}-\frac54\nonumber\\
&+&{5\lambda\over 8\pi^2}\left[ 2\left(\omega\,{\rm artanh}
{1\over \omega}\right)^2
-\omega\,{\rm artanh}{1\over \omega}-
{\omega^2\over {\omega^2-1}} \right] \nonumber\\
&+&{5\lambda^{3/2}\over 8\pi^3} \biggl[
3\left({\omega^2-1}-\omega\sqrt{\omega^2-1}\right)\left(\omega\,{\rm artanh}
{1\over \omega}\right)^2 \nonumber\\
&&+6 \left( \omega\sqrt{\omega^2-1}-\omega^2-{\omega\over \sqrt{\omega^2-1}}
\right)\omega
\,{\rm artanh}{1\over \omega}\nonumber\\
&&+{\omega\over ({\omega^2-1})^{3/2}}+3{\omega^2\over {\omega^2-1}}+
6{\omega\over \sqrt{\omega^2-1}}-3\omega\sqrt{\omega^2-1}+3\omega^2
\biggr]
\end{eqnarray}
and similarly complicated expressions arise for $B$ and $C$, which
in the collisionless limit were pure numbers.

The Fourier transform of this expression determines the kernel
in the convolution integral of \gl{scp}. At order $\lambda^1$,
it can still be expressed in terms of well-known special functions
\cite{Rapid}, whereas at order $\lambda^{3/2}$ this would involve
rather intractable integrals over Lommel functions. However, all
that is needed for finding analytical solutions is their power
series representations which are comparatively simple.
Given them, it is as easy as before to solve the perturbation equations,
however one finds that the asymptotic behaviour $x\gg1$ eventually
becomes sensitive to higher and higher loop orders. The reason for
this is that higher loop orders come with increasingly singular
contributions at $\omega=\pm1$ to $A(\omega)$, and the large-$x$
behaviour is dominated by the latter. This could be cured by
a further resummation similar to the one introduced for
hot quantum chromodynamics in Ref.~\cite{FR}, but it turns out
that a particular Pad\'e-approximant based on the perturbative result
reflects the effects of this further resummation quite well \cite{Nrs3}.
The result for the density perturbations in a scalar plasma with
$\lambda\phi^4$-interactions and $\lambda=1$ are shown in Fig.~\ref{plot}
by the dashed line,
where it is compared with the collisionless case (full line) and the one of
a perfect radiation fluid (dotted line).
The effects of the self-interactions within the ultrarelativistic plasma
become important only for $x\gtrsim \pi$, where the strong collisionless
damping is somewhat reduced and the phase velocity is smaller than 1.

\begin{figure}
\centerline{ \epsfxsize=5in
\epsfbox{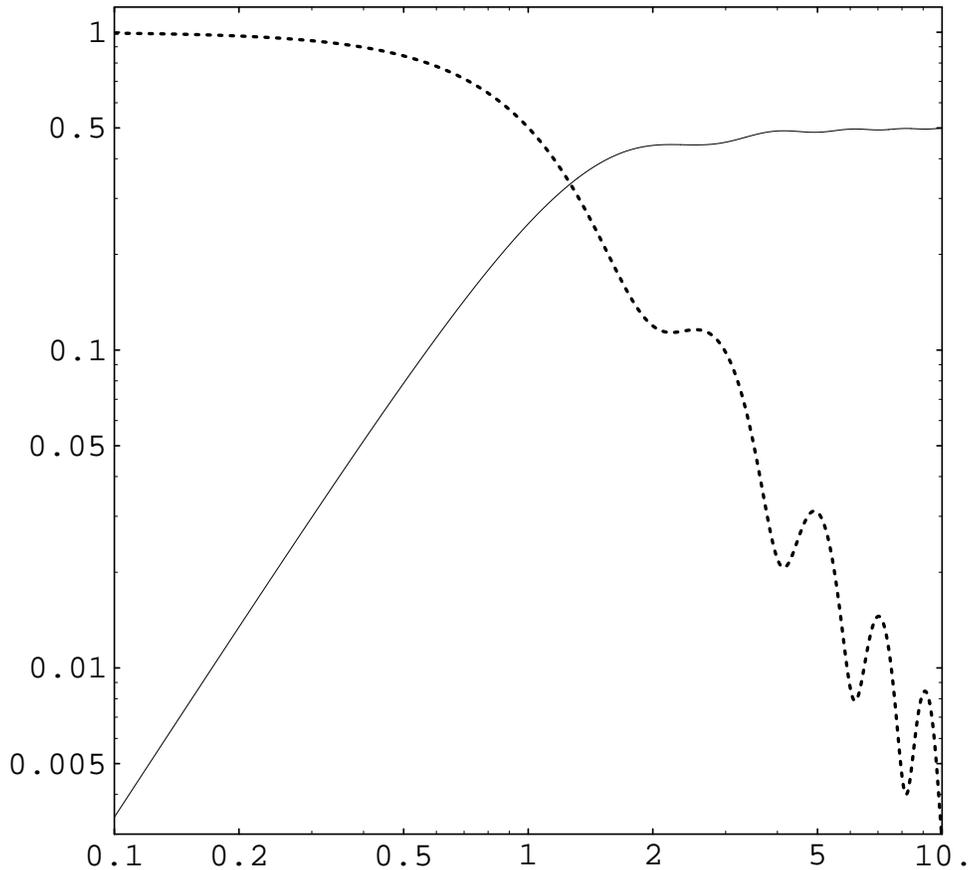} }
\caption{Rotational perturbations in a two-component system consisting
of 50 \% perfect radiation fluid and 50 \% ultrarelativistic scalar plasma
($\lambda=1$).
Given is the velocity amplitude for the total system (full line)
and the weakly interacting plasma (dashed line) in arbitary units
as a function of $x/\pi$.}
\label{v2}\end{figure}

A full analysis of scalar, vector, and tensor perturbations
in the general case of a two-component system containing also
a perfect radiation fluid is given in Ref.~\cite{Nrs3}.
Let us just mention one of the more spectacular results, which
arise in the case of vector (rotational) perturbations.
This case has not been investigated much previously, presumably
because in the perfect-fluid case there are no regular solutions ---
rotational perturbations necessarily lead to strongly anisotropic
initial singularities. This can be explained by the Helmholtz-Kelvin
circulation theorem \cite{HKth} which states that in a perfect fluid
the circulation around a closed curve following the motion of matter
is conserved. However, this theorem does not apply generally.
Indeed, in a (nearly) collisionless medium
one can have small initial anisotropies in the distribution
function that gives rise to a growing
vorticity on superhorizon scales \cite{Rebhan92},
which decays after horizon crossing through directional dispersion.
In a two-component system one can even have such perturbations which
do not decay by arranging for vorticity in a perfect-fluid component
that is compensated by an initially matching one with reversed sign in the
nearly collisionless plasma component. Then net vorticity is
generated by the decay of the vector perturbation in the plasma
component, see Fig.~\ref{v2}. This is particularly interesting
in that vector perturbations generally lead to the generation of
primordial magnetic fields at the time when the universe
changes from radiation to matter domination \cite{RApJ}.
Because tiny primordial magnetic fields can act as seed fields for
galactic dynamos, such rotational perturbations may therefore be of interest
with respect to the still unsolved problem of the origin of
galactic and intergalactic magnetic fields.

\acknowledgments

I am grateful to Ulli Kraemmer, Herbert Nachbagauer, and Dominik Schwarz,
for their most enjoyable collaboration in various stages of the presented
work.

\begin{table}
\begin{tabular}{lcc}
$T^{\alpha\beta\mu\nu}_1=\eta ^{\alpha \nu }\,\eta ^{\beta \mu }+
\eta ^{\alpha \mu }\,\eta ^{\beta \nu }$  \\
$T^{\alpha\beta\mu\nu}_2=u^\mu \,\left( u^\beta
\,\eta ^{\alpha \nu }+u^\alpha \,\eta ^{\beta \nu }\right) +
u^\nu \,\left( u^\beta \,\eta ^{\alpha \mu }
+u^\alpha \,\eta ^{\beta \mu }
\right) $ \\
$T^{\alpha\beta\mu\nu}_3=u^\alpha \,u^\beta \,u^\mu \,u^\nu $ \\
$T^{\alpha\beta\mu\nu}_4=\eta ^{\alpha \beta }\,\eta ^{\mu \nu }$  \\
$T^{\alpha\beta\mu\nu}_5=u^\mu \,u^\nu \,\eta ^{\alpha \beta }+
u^\alpha \,u^\beta \,\eta ^{\mu \nu }$ \\
$T^{\alpha\beta\mu\nu}_6=u^\beta \,\left( \bar K^\nu \,
\eta ^{\alpha \mu }+\bar K^\mu \,\eta ^{\alpha \nu }\right) +
\bar K^\beta \,\left( u^\nu \,\eta ^{\alpha \mu }
+u^\mu \,\eta ^{\alpha \nu }
\right) $ \\
$\;\;\;\;\;\;\;\;\,\,\,\,\,\;\;\;\;\;\;\;\;\;\;\;
+\;u^\alpha \,\left( \bar K^\nu \,\eta ^{\beta \mu }+
\bar K^\mu \,\eta ^{\beta \nu }\right) +\bar K^\alpha \,\left( u^\nu
\,\eta ^{\beta \mu }+u^\mu \,\eta ^{\beta \nu }\right) $ \\
$T^{\alpha\beta\mu\nu}_7=\bar K^\nu \,u^\alpha \,u^\beta \,u^\mu +
\bar K^\mu \,u^\alpha \,u^\beta \,u^\nu +
\bar K^\beta \,u^\alpha \,u^\mu \,u^\nu +
\bar K^\alpha \,u^\beta \,u^\mu \,u^\nu $ \\
$T^{\alpha\beta\mu\nu}_8=
\bar K^\beta \,\bar K^\nu \,\eta ^{\alpha \mu }+
\bar K^\beta \,\bar K^\mu \,\eta ^{\alpha \nu }+
\bar K^\alpha \,\bar K^\nu \,\eta ^{\beta \mu }+
\bar K^\alpha \,\bar K^\mu \,\eta ^{\beta \nu }$ \\
$T^{\alpha\beta\mu\nu}_9=
\bar K^\mu \,\bar K^\nu \,u^\alpha \,u^\beta +
\bar K^\alpha \,\bar K^\beta \,u^\mu \,u^\nu $ \\
$T^{\alpha\beta\mu\nu}_{10}=\left( \bar K^\beta \,u^\alpha +
\bar K^\alpha \,u^\beta \right) \,
\left( \bar K^\nu \,u^\mu +\bar K^\mu \,u^\nu \right) $ \\
$T^{\alpha\beta\mu\nu}_{11}=
\bar K^\beta \,\bar K^\mu \,\bar K^\nu \,u^\alpha +
\bar K^\alpha \,\bar K^\mu \,\bar K^\nu \,u^\beta +
\bar K^\alpha \,\bar K^\beta \,\bar K^\nu \,u^\mu +
\bar K^\alpha \,\bar K^\beta \,\bar K^\mu \,u^\nu $ \\
$T^{\alpha\beta\mu\nu}_{12}=\bar K^\alpha \,\bar K^\beta \,
\bar K^\mu \,\bar K^\nu $ \\
$T^{\alpha\beta\mu\nu}_{13}=
\bar K^\mu \,\bar K^\nu \,\eta ^{\alpha \beta }+
\bar K^\alpha \,\bar K^\beta \,\eta ^{\mu \nu }$ \\
$T^{\alpha\beta\mu\nu}_{14}=
\left( \bar K^\nu \,u^\mu +\bar K^\mu \,u^\nu
\right) \,\eta^{\alpha \beta }+\left( \bar K^\beta \,u^\alpha +
\bar K^\alpha \,u^\beta \right) \,\eta^{\mu \nu }$ \\
\end{tabular}
\vskip 0.5cm

\caption{A basis of 14 independent tensors
$T_i^{\alpha\beta\mu\nu}$ built from $\eta^{\mu\nu}$,
$u^\mu=\delta^\mu_0$,
and $\bar K^\mu\equiv K^\mu/k=(\omega,{\bf k}/k)$.
\label{table1}}
\end{table}

\begin{table}
\begin{tabular}{lcc}
$c_1 = \,{ {{1}\over 8}{\bar K^4}A +
   {1\over 2}{\bar K^2}B + {{1}\over 4}C +
   {{1}\over {32}}{\bar K^4} +
   {{11}\over {24}}\bar K^2 + {{{1}}\over 6} }$ \\
$c_2 = \,{ {{5}\over 8}{\bar K^6}A + {\bar K^4}B +
   {1\over 4}{\bar K^2}C +
   {{5}\over {32}}{\bar K^6} +
   {{19}\over {24}}{\bar K^4} +
   {1\over {12}}{\bar K^2} - {{{1}}\over 3}}$\\
$c_3 = {\bar K^2}\left\{ \,{ {{35}\over 8}{\bar K^6}A +
   {{5}\over 2}{\bar K^4}B +
   {{1}\over 4}{\bar K^2}C +
   {{35}\over {32}}{\bar K^6} +
   {{25}\over {24}}{\bar K^4} +
   {{7}\over {12}}{\bar K^2} - {1\over 3}} \right\}$\\
$c_4 = {\bar K^2}\left\{
        \,{ {{1}\over 8}{\bar K^2}A - {1\over 2}B -
   {{1}\over 4}C + {{1}\over {32}}{\bar K^2} -
   {{13}\over {24}}} \right\}$\\
$c_5 = {\bar K^2}\left\{ \,{ {{5}\over 8}{\bar K^4}A -
   {{1}\over 2}{\bar K^2}B -
   {1\over 4}C +
   {{5}\over {32}}{\bar K^4} -
   {{17}\over {24}}{\bar K^2} +
   {1\over {12}}} \right\}$\\
$c_6 = \omega\left\{ \,{ {{-5}\over 8}{\bar K^4}A -
   \bar K^2B - {{1}\over 4}C -
   {{5}\over {32}}{\bar K^4} -
   {{19}\over {24}}\bar K^2 -
   {{{1}}\over {12}}} \right\}$\\
$c_7 = \omega\left\{ \,{ {{-35}\over 8}{\bar K^6}A -
   {{5}\over 2}{\bar K^4}B -
   {1\over 4}{\bar K^2}C -
   {{35}\over {32}}{\bar K^6} -
   {{25}\over {24}}{\bar K^4} -
   {{7}\over {12}}\bar K^2 +
   {{{1}}\over 3}}\right\}$\\
$c_8 = \,{ \left({{5}\over 8}{\bar K^2} +
   {1\over 2}\right){\bar K^2}A + \left( \bar K^2 +
   {{1}\over 2}\right)B +  {{1}\over 4}C +
   {{5}\over {32}}{\bar K^4} +
   {{11}\over {12}}\bar K^2 + {{5}\over {12}}}$\\
$c_9 = \,{ \left({{35}\over 8}{\bar K^2} +
   {{15}\over 4}\right){\bar K^4}A +
   \left({{5}\over 2}{\bar K^2} + 3\right)\bar K^2B +
   \left({1\over 4}{\bar K^2} + {{1}\over 2}\right)C
+
   {{35}\over {32}}{\bar K^6} +
   {{95}\over {48}}{\bar K^4} +
   {{7}\over 3}\bar K^2 + {{{1}}\over 6}} $\\
$c_{10} = \,{ \left({{35}\over 8}{\bar K^2} +
   {{15}\over 4}\right){\bar K^4}A +
   \left({{5}\over 2}{\bar K^2} +
   {{3}\over 2}\right)\bar K^2B +
   {1\over 4}{\bar K^2}C
+
   {{35}\over {32}}{\bar K^6} +
   {{95}\over {48}}{\bar K^4} +
   {{5}\over 6}\bar K^2 + {{{1}}\over 6}} $\\
$c_{11} = \omega\left\{ \,{ \left({{-35}\over 8}{\bar K^2} -
   {{5}\over 2}\right)\bar K^2A -
   \left({{5}\over 2}\bar K^2 + 1\right)B -
   {{1}\over 4}C -
   {{35}\over {32}}{\bar K^4} -
   {{5}\over 3}\bar K^2 -
   {{3}\over 4}}\right\}$\\
$c_{12} = \,{ \left({{35}\over 8}{\bar K^4} + 5\bar K^2 +
   1\right)A + \left({{5}\over 2}\bar K^2 + 2\right)B +
   {{1}\over 4}C +
   {{35}\over {32}}{\bar K^4} +
   {{55}\over {24}}\bar K^2 + {{7}\over 6}}\qquad$\\
$c_{13} = \,{ \left({{5}\over 8}{\bar K^2} +
   {1\over 2}\right){\bar K^2}A - {1\over 2}{\bar K^2}B -
   {{1}\over 4}C + {{5}\over {32}}{\bar K^4} -
   {{7}\over {12}}\bar K^2 - {{{1}}\over {12}}}$\\
$c_{14} = \omega\left\{ \,{ {{-5}\over 8}{\bar K^4}A +
   {1\over 2}{\bar K^2}B +
   {{1}\over 4}C -
   {{5}\over {32}}{\bar K^4} +
   {{17}\over {24}}\bar K^2 -
   {{{1}}\over {12}} }\right\}$ \\
   \end{tabular}
\vskip 0.5cm
\caption{The structure of the conformally covariant
gravitational polarization tensor
$\tilde\Pi^{\mu\nu\alpha\beta}/\rho =  \sum_{i=1}^{14} c_i
T_i^{\mu\nu\alpha\beta}$ in terms of
$A\equiv\tilde\Pi_{0000}/\rho$,
$B\equiv\tilde\Pi_{0\mu}{}^\mu{}_0/\rho$, and
$C\equiv\tilde\Pi_{\mu\nu}{}^{\mu\nu}/\rho$.
\label{table2}}
\end{table}


\begin{references}

\bibitem{Boerner}G. B\"orner, {\it The Early Universe: Facts and Fiction}
(Springer-Verlag, Berlin, 1993).
\bibitem{Smoot}G. F. Smoot et al., Astrophys. J. {\bf 396}, L1 (1992).
\bibitem{Lifshitz}E. Lifshitz, Zh. Eksp. Teor. Fiz. {\bf 16}, 587 (1946);\\
         E. Lifshitz and I. Khalatnikov, Adv. Phys. {\bf 12}, 185 (1963).
\bibitem{Bardeen}J. M. Bardeen, Phys. Rev. D {\bf 22}, 1882 (1980).
\bibitem{EB}G. F. R. Ellis and M. Bruni, Phys. Rev. D {\bf 40}, 1804 (1989);\\
G. F. R. Ellis, J. Hwang and M. Bruni, Phys. Rev. D {\bf 40}, 1819 (1989).
\bibitem{Kin}J. Ehlers, in {\it General Relativity and Gravitation}, edited
         by R. K. Sachs (Academic Press, New York, 1971);\\ J. M. Stewart,
         {\it Non-equilibrium Relativistic Kinetic Theory}, (Springer-Verlag,
         New York, 1971).
\bibitem{Kapusta}J. I. Kapusta, {\em Finite-temperature field theory},
         (Cambridge University Press, Cambridge, 1989);\\
M. Le Bellac, {\em Thermal Field Theory} to appear in
Cambridge University Press.
\bibitem{KR}U. Kraemmer and A. Rebhan, Phys. Rev. Lett. {\bf 67},
         793 (1991).
\bibitem{Rapid}H. Nachbagauer, A. K. Rebhan, and D. J. Schwarz,
Phys. Rev. D {\bf 51}, R2504 (1995).
\bibitem{Rebhan91}A. Rebhan, Nucl. Phys. {\bf B351}, 706 (1991).
\bibitem{ABFT}J. Frenkel and J. C.
         Taylor, Z. Phys. C {\bf49}, 515 (1991);\\
	F. T. Brandt, J. Frenkel and J. C.
         Taylor, Nucl. Phys. {\bf B374}, 169 (1992);\\
	A. P. de Almeida, F. T.  Brandt and J. Frenkel,
         Phys. Rev. D {\bf 49}, 4196 (1994);\\
	J. Frenkel, E. A. Gaffney and J. C. Taylor,
         Nucl. Phys. {\bf B439}, 131 (1995).
\bibitem{AKRDS} A. K. Rebhan and D. J. Schwarz,  Phys. Rev. D
          {\bf 50}, 2541 (1994).
\bibitem{Rebhan92}A. Rebhan, Nucl. Phys. {\bf B368}, 479 (1992).
\bibitem{Bond}J. R. Bond and A. S. Szalay,
Astrophys. J. {\bf 274}, 443 (1983).
\bibitem{Nrs2}H. Nachbagauer, A. K. Rebhan, and D. J. Schwarz,
preprint DESY 95-136, to appear in Phys. Rev. D (1995).
\bibitem{BP}E. Braaten and R. D. Pisarski,
Nucl. Phys. {\bf B337}, 569 (1990).
\bibitem{FR}U. Kraemmer, A. K. Rebhan and H. Schulz, Ann. Phys. (NY)
{bf 238}, 286 (1995);\\
F. Flechsig and A. K. Rebhan, preprint DESY 95-170.
\bibitem{Nrs3}H. Nachbagauer, A. K. Rebhan, and D. J. Schwarz,
preprint DESY 95-191.
\bibitem{HKth}B. J. T. Jones, Rev. Mod. Phys. {\bf 48}, 107 (1976).
\bibitem{RApJ}A. Rebhan, Astrophys. J. {\bf392}, 385 (1992).

\end{references}
\end{document}